\documentclass[12pt]{article}
\usepackage{graphicx}
\usepackage{psfrag}
\usepackage{epstopdf}

\begin{document}

\begin{titlepage}
\null\vspace{-62pt}

\pagestyle{empty}
\begin{center}

\vspace{1.0truein} {\Large\bf A proposal for the Yang-Mills vacuum and mass gap }

\vspace{1in}
{\large Dimitrios Metaxas} \\
\vskip .4in
{\it Department of Physics,\\
National Technical University of Athens,\\
Zografou Campus, 15780 Athens, Greece\\
metaxas@central.ntua.gr}\\

\vspace{0.5in}

\vspace{.5in}
\centerline{\bf Abstract}

\baselineskip 18pt
\end{center}

\noindent

I examine a set of Feynman rules, and the resulting effective action, that were proposed in order to incorporate the constraint of Gauss's law
in the perturbation expansion of gauge field theories. 
A set of solutions for the Lagrangian and Hamiltonian equations of motion in Minkowski space-time,
as well as  their stability, are investigated.
A discussion of the Euclidean action, confinement, and the strong-CP problem is also included.
The properties and symmetries of the perturbative and the confining vacuum are explored, as well as the possible
transitions between them, and the relations with phenomenological models of the strong interactions.

\end{titlepage}

\newpage
\pagestyle{plain}
\setcounter{page}{1}
\newpage

\section{Introduction}

In a previous work \cite{dkm}, I considered the possibility of expressing the constraint of Gauss's law
in the perturbative expansion of gauge field theories  via a Lagrange multiplier
field, $\lambda$, and argued for the generation
of an effective potential term of the Coleman-Weinberg type \cite{CW} for $\lambda$,
and its relation to the problems of the mass gap and confinement
in the non-Abelian case.

Here, I elaborate on the  consequences of the procedure and proposed effective action, I derive an effective Hamiltonian, and
examine the energy and stability of solutions
with ``bubbles'' of the chromoelectric field.
A discussion of the Euclidean action, the vacua and possible related vacuum transitions, confinement, and the strong-CP problem is also included.
The symmetries and other properties of the perturbative and the confining vacuum are explored, and connections are made with older phenomenological models of the strong interactions.

In particular, the works of \cite{kogut} can be mentioned, where some interesting and intuitive phenomenological models of the confining mechanism have been proposed, with the addition of a scalar field and an associated effective potential term, that modify the dielectric and fermion condensate properties of the theory at its minimum.
The present work also includes a scalar field, the Lagrange multiplier, and an associated effective potential term. Here, however, $\lambda$ has no kinetic term and no additional degrees of freedom, hence there is no symmetry breaking, and the effective potential term appears ``inverted'', with the opposite sign.

Although this effective potential term here is unbounded below, because of the interplay of the gauge kinetic and gradient terms, as well as the constraint of Gauss's law, stability is proven for all classical solutions.
Also, two vacua emerge, a local minimum (the perturbative, Coulomb vacuum, $\Omega_0$, at $\lambda=0$) and a maximum of the effective potential (the confining vacuum, $\Omega_\mu$, at the generated mass scale $\lambda^2=\mu^2$), that are also shown to be quantum mechanically stable; there are no finite action Euclidean solutions that mediate their decay.
There are, however, stable Lorentzian solutions, ``bubbles'' of the chromoelectric field, ``glueballs'', that connect the two vacua and  can mediate the transitions between them. They are solitonic solutions with a finite mass ``gap'' of order $\mu /g^2$ (where $g$ is the coupling constant of the non-Abelian theory).

Once the vacuum structure of the theory is better understood, several properties of the Yang-Mills theory, that were also expected to be related, can be easily seen: confinement \cite{conf}, bag model \cite{bag}, chiral symmetry breaking \cite{chiral}, as well as a possible solution to the strong-CP problem \cite{cp}.

As far as the Lorentz invariance of the theory is concerned,
the two vacua admit a Lorentz invariant energy-momentum tensor, the one expected by the bag model, but they are completely stable for pure Yang-Mills theory at zero temperature, and there is no Lorentz invariant energy-momentum tensor that connects them (at least not with the effective action derived in this work, which concerns pure Yang-mills at zero temperature). Any transitions between the two vacua happen in non-trivial backgrounds of finite temperature or fermion density.

Although the work presented here can be related to older phenomenological models, and can also be considered as an effective action that describes the properties of the strong interactions, it should be stressed that it is derived from first principles, namely the treatment of the constraints in the quantum theory, and is proposed as a complete and exact description of the Yang-Mills vacuum and associated features.

The layout of this work is as follows:
in Sec.~2, I start with a description of the combinatorics for the Abelian case
in order to explain the procedure in a simpler setting, but also to show that the method proposed here does not change the perturbative behavior of the theory.
In Sec.~3, I consider
the non-Abelian, self-interacting case, show the derivation of an effective potential term for 
$\lambda$, and examine the solitonic, ``bubble'' solutions, and their stability in the Lagrangian and Hamiltonian frameworks in Minkowski space-time.
In Sec.~4, I give a preliminary discussion of the Euclidean action, the vacua and possible solutions.
In Sec.~5, the previous considerations are utilized in order to obtain a better picture of the  confining vacuum and mechanism.
In Sec.~6, I give a more detailed treatment of the Euclidean action and show that there are no finite action solutions that mediate vacuum decay of either the perturbative or the confining vacuum of pure Yang-Mills theory at zero temperature.
The only solutions of the Euclidean equations are the usual Yang-Mills instantons,  which exist at both the perturbative and the confining vacuum.
In Sec.~7, I examine the symmetry properties of the theory at the two vacua, discuss Lorentz invariance, the bag model, and chiral symmetry breaking.
In Sec.~8, I give some arguments towards the resolution of the strong-CP problem.
In Sec.~9, I conclude with some comments.

\section{The Abelian theory}

In order to investigate the consequences of the constraint of Gauss's law in the perturbation expansion
of gauge field theories I will start with the Abelian case, including a massive fermion, with Lagrangian $\cal L$ and action
\begin{equation}
S=\int
{\cal L}= \int  -\frac{1}{4}F_{\mu\nu}^2 + \bar{\psi} (i \gamma^{\mu}D_{\mu} - m) \psi \,,
\end{equation}
where $F_{\mu\nu}=\partial_{\mu} A_{\nu} - \partial_{\nu} A_{\mu}$ and $D_{\mu}=\partial_{\mu} + i e A_{\mu}$.

Integrations are over $d^4 x$ and
the metric conventions are $g_{\mu\nu}=(+---)$, $\partial_{\mu}=(\partial_0, \partial_i)$,
$\partial^{\mu}=(\partial_0, -\partial_i)$.
Generally,
$r$ will denote the three-dimensional, spatial distance.

Since the Lagrangian is independent of $\dot{A_0}=\partial_0 A_0$, the
respective equation of motion for that field, namely
\begin{equation}
\frac{\delta S}{\delta A_0}=0,
\label{gauss1}
\end{equation}
is not a dynamical equation, but, rather, a constraint corresponding to Gauss's law,
which can be incorporated in the perturbative expansion via a Lagrange multiplier field,
$\lambda$, in the path integral
\begin{equation}
Z(J_{\mu}, \Lambda)=\int[dA_{\mu}][d\psi][d\bar{\psi}][d\lambda] e^{i \int\tilde{\cal L}},
\end{equation}
where
\begin{eqnarray}
\nonumber
  \tilde{\cal L} &=&\frac{1}{2}(\partial_0 A_i - \partial_i A_0)^2
                   -\frac{1}{4} F_{ij}^2\\ \nonumber
    &+& \bar{\psi} (i \gamma^{\mu}\partial_{\mu} -e\gamma^0 A_0 +e \gamma^i A_i - m) \psi \\ \nonumber
    &-& \lambda(\nabla^2 A_0-\partial_0\partial_i A_i + e \bar{\psi}\gamma^0\psi) \\
    &-& \frac{1}{2} (\partial_0 A_0 -\partial_i A_i +\partial_0 \lambda)^2
                    +A_0 J_0 - A_i J_i + \lambda \Lambda.
\label{l1}
\end{eqnarray}
In the above equation the first and the second lines contain the original gauge and fermion terms,
the third line is the constraint $\lambda \frac{\delta S}{\delta A_0}$, implemented with
a gauge-invariant $\lambda$, and the last line contains
the gauge-fixing term and the sources $J_{\mu}, \Lambda$.
A special gauge-fixing condition was used, since the associated term, which can be derived by the
usual Faddeev-Popov procedure, gives the simplest set of Feynman rules. 
Other gauge conditions can be used \cite{dkm}, with a similar combinatoric result as described below:

After the usual inversion procedures one obtains the propagators with momentum $k$,
\begin{eqnarray}
  G_{00} &=& -\frac{1}{k^2}-\frac{1}{\vec{k}^2} \\
  G_{\lambda\lambda} &=& -\frac{1}{\vec{k}^2}  \\
  G_{0\lambda} &=& \frac{1}{\vec{k}^2} \,=\, G_{\lambda 0}\\
  G_{ii} &=& \frac{1}{k^2}.
\label{props}
\end{eqnarray}
One can easily deduce the vertices from (\ref{l1}), and observe the fact that
the propagators are combined in all interactions so as to reproduce
all the usual QED diagrams. $G_{00}$, $G_{\lambda\lambda}$ and $G_{0\lambda}$
appear together and their sum gives the ordinary $0-0$ propagator in Feynman gauge.
For example, for two static current sources
separated by a spatial distance, $\vec{r}$,
one obtains the usual Coulomb
interaction energy
 from the sum of the diagrams in Fig.~1
in the static limit of $k_0=0$,
\begin{equation}
V_{\rm Coul}(r) =4\pi\alpha_e \int \frac{d^3 {k}}{(2 \pi)^3} \frac{e^{i\vec{k}\cdot\vec{r}}}{\vec{k}^2}
= \frac{\alpha_e}{r},
\label{coulomb}
\end{equation}
with $\alpha_e = \frac{e^2}{4\pi}$.

\section{The non-Abelian theory}

I now consider the case of the non-Abelian, Yang-Mills gauge theory, with coupling $g$, and  gauge group $G$, with generators $T^a$ and structure constants $f^{abc}$, and initial action
\begin{equation}
S_0=\int -\frac{1}{4}F^a_{\mu\nu} F^{a\mu\nu}  \,,
\end{equation}
with $F^a_{\mu\nu}=\partial_{\mu}A^a_{\nu}-\partial_{\nu}A^a_{\mu} - g f^{abc}A^b_{\mu}A^c_{\nu}$.

The theory is gauge invariant, with
$A_{\mu}\rightarrow\omega A_{\mu} \omega^{-1} + \frac{i}{g}\omega\partial_{\mu} \omega^{-1}$
 under the local gauge transformation
$\omega(\alpha) = e^{iT^a\alpha^a (x)} \in G$
(with the usual notation $A_{\mu}= T^a A^a_{\mu}$).
Although the addition of fermions will not be considered here, a massive fermion in the representation $R$
can be included with the term $ \bar{\psi} (i \gamma^{\mu}D_{\mu} - m) \psi$ in the Lagrangian
 with $\psi\rightarrow\omega_R(\alpha)\psi$,
and $D_{\mu}=\partial_\mu + i g\,A_{\mu}^a \, T_R^a$.

After imposing the constraint in
\begin{equation}
\tilde{S} = S_0 + \int \lambda^a \frac{\delta S_0}{\delta A_0},
\label{arg1}
\end{equation}
the theory is still gauge-invariant with $\lambda\rightarrow\omega\lambda\omega^{-1}$, and can be gauge-fixed 
similarly to the Abelian case.
The resulting gauge field propagators
are the same as the Abelian theory and diagonal in color indices. Other gauge conditions are possible \cite{dkm}, the general result being, as described before, the missing Coulomb interaction, and its reconstruction with the modified Feynman rules.

The incorporation of the constraint of Gauss's law via the term
$\lambda^a \frac{\delta S_0}{\delta A^a_0}$ has the additional effect
of introducing interactions between the gauge field and $\lambda$. These are
the same as the usual interactions, with one $A_0$ leg replaced by $\lambda$.
For example, in Fig.~2, a vertex of the non-Abelian theory is shown
together with the new corresponding vertex with the same value.

The usual QCD interactions can be reproduced, with the exception
that, for diagrams with external $\lambda$ legs, the Coulomb interaction is
missing in the internal propagators: in Fig.~3, this is shown for the $A_i-A_j$ propagator, with momentum $k$,
and external, constant $\lambda$ fields, where the missing Coulomb interaction
gives a factor of $g^2 C_2 \lambda^2 \frac{k_i k_j}{\vec{k}^2}$,
where $\lambda^2=\lambda^a\lambda^a$ and $f^{acd}f^{bcd}=C_2 \delta^{ab}$.
This amounts to a mass term
 in loops like Fig.~4, where the
$\lambda-A_0$ and $\lambda-\lambda$ interaction cannot be inserted in the loop,
and has the effect of generating a gauge invariant
effective potential from these terms \cite{dkm},
which would otherwise add up to zero.
It is of the Coleman-Weinberg form \cite{CW},
\begin{equation}
U(\lambda)= \frac{(\alpha_s C_2)^2}{4}\lambda^4 \left(\ln\frac{\lambda^2}{\mu^2}-\frac{1}{2}\right),
\label{m0}
\end{equation}
with $\alpha_s = g^2/4\pi$,
renormalized at a scale $\mu$ where $dU/d\lambda=0$,
and appears in the effective action with the opposite sign (it is upside-down).

The resulting, gauge invariant, effective action in Minkowski space,
\begin{equation}
S_{\rm M, eff}= \int -\frac{1}{4}F^a_{\mu\nu} \, F^{a \,\mu\nu} +
                        \lambda^a \, D_i  F^{a i0} + U(\lambda),
\label{meff1}
\end{equation}
can also be written in terms of the (chromo)-electric and -magnetic fields,
\begin{equation}
E^a_i = F^a_{0i}=F^{a i0},\,\,\, B^a_i=-\frac{1}{2}\epsilon^{ijk}F^a_{jk},
\end{equation}
 as
\begin{equation}
S_{\rm  eff}= \int \frac{1}{2}{E_i^a}E_i^a -\frac{1}{2}{B_i^a}B_i^a +
                        \lambda^a\, {D_i} {E_i^a} + U(\lambda),
\label{meff2}
\end{equation}
and the variational equations become
\begin{equation}
\frac{\delta}{\delta\lambda^a}=0 \,\Rightarrow
D_i E_i^a = -\frac{\partial U}{\partial \lambda^a},
\label{m1}
\end{equation}
\begin{equation}
\frac{\delta}{\delta A_0^a} =0 \,\Rightarrow
D_i^2 \lambda^a = D_i E_i^a,
\label{m2}
\end{equation}
\begin{equation}
\frac{\delta}{\delta A_i^a} =0 \Rightarrow
D_0 \, E_i^a =  (D\times B)_i^a + D_0 D_i \lambda^a.
\label{m3}
\end{equation}
Generally, their solution requires a choice of gauge, which can be imposed at this level, or added in the effective action as usual. However, one can obtain a set of solutions by inspection,
setting $E_i^a = D_i \lambda ^a$, and demanding $D_i^2 \lambda^a =-\frac{\partial U}{\partial \lambda^a}$, further setting $A_i^a=0, B_i^a=0$, 
hence $E_i =-\partial_i A_0 = \partial_i \lambda$, and considering the
solutions of the equation
\begin{equation}
\nabla^2 \lambda^a =-\frac{\partial U}{\partial \lambda^a}.
\label{m4}
\end{equation}
This is the same as the equation that would describe tunneling in a three-dimensional model of a scalar field with an inverted Coleman-Weinberg potential term. There, with a potential unbounded from below, the theory would develop an instability; here, this is not clear yet, since there is no kinetic term for $\lambda$ in the effective action. The solutions of (\ref{m4}) are spherically symmetric ``bubbles'' of non-zero $\lambda^a$, with
$\lambda^a \approx \mu$ near the center, and finite radius $R\approx \frac{1} {\alpha_s \, C_2 \, \mu}$ 
(the color index, $a$, with the non-zero field values, is arbitrary and can be rotated by a gauge transformation).

Obviously, there is also the solution with $\lambda =0$ and the other fields zero everywhere, which corresponds to the usual, perturbative Yang-Mills vacuum.
The solution with $\lambda^2=\mu^2$, equal everywhere to the other, non-zero extremum of $U$, will be discussed later.
There are also solutions that consist of combinations of various bubbles, separated at finite distances from each other, and with varying relative gauge orientations.

The covariance of (\ref{m1}-\ref{m3}), as well as the invariance of the generated effective potential, implies that we can apply time-independent gauge transformations, $\omega$, to the solutions with zero $A_i$, to transform to a pure gauge $A_i= \frac{i}{g} \omega^{-1} \partial_i \,\omega$, while keeping the magnetic field equal to zero, and satisfying the remaining requirements, $E_i^a = D_i \lambda ^a\,$, $D_i^2 \lambda^a =-\frac{\partial U}{\partial \lambda^a}$, for the bubble solutions,  thus these also exist in the well-known, topologically non-trivial, sectors of the Yang-Mills theory \cite{cw}.

In order to proceed to a canonical formalism, one can exploit the gauge invariance of (\ref{meff1}) to set $A_0^a=0$, 
hence $E_i^a = \dot{A_i^a}$, and consider the effective action
\begin{equation}
S_{\rm eff, 0}=\int L_{\rm eff, 0} = \int \frac{1}{2} \dot{A_i^a}\dot{A_i^a} - \frac{1}{2} B_i^a B_i^a
                        + \lambda^a D_i \dot{A_i^a} +U,
\label{meff2}
\end{equation}
from which the equivalent set of equations
\begin{equation}
\frac{\delta}{\delta \lambda^a}=0 \Rightarrow D_i \dot{A_i^a}= D_i E_i^a = -\frac{\partial U}{\partial \lambda^a},
\label{m11}
\end{equation}
\begin{equation}
\frac{\delta}{\delta A_i^a}=0 \Rightarrow \partial_0^2 A_i^a = \dot{E_i^a}=
                                       (\nabla\times {B})^a_i +\partial_0 D_i \lambda^a,
\label{m22}
\end{equation}
are derived, that also admit the bubble solutions
with $E_i^a =D_i \lambda^a,\, B_i^a =0,\, D_i^2 \lambda^a = -\partial U/ \partial\lambda^a$.

Specifically, there is a gauge transformation, $\tilde{\omega}$,  that changes the previous
solution $A_i^a=0, \, A_0^a =-\lambda^a, \, E_i^a =\partial_i \lambda^a$ to this gauge with
$\tilde{A_0^a}=0, \,  \tilde{A_i}=\frac{i}{g} \tilde\omega^{-1} \partial_i \,\tilde\omega , \,
\tilde{E_i} =\tilde{\omega} E_i \tilde\omega^{-1}$, and still allows the freedom of time-independent gauge transformations.

Now, the three canonical variables, $Q^a_i = A^a_i$, are further constrained by (\ref{m11}), and admit
the conjugate canonical momenta, 
\begin{equation}
P_i^a = \frac{\partial L_{\rm eff,  0}}{\partial \dot{A_i^a}} = \dot{A_i^a} -D_i\lambda^a = E_i^a-D_i\lambda^a .
\label{pq}
\end{equation}
An effective Hamiltonian can be defined as
\begin{eqnarray}
H_{\rm eff} &=&\int_3 P_i^a \dot{Q_i^a} - L_{\rm eff, 0} = \nonumber \\
                   &=&\int_3 \frac{1}{2}P_i^a P_i^a + \frac{1}{2} B_i^a B_i^a +
                 \frac{1}{2} (D_i \lambda^a)^2 + P_i^a D_i\lambda^a - U,
\label{heff}
\end{eqnarray}
where $\int_3$ denotes integration in three-dimensional space.
The canonical equations
\begin{equation}
\dot{Q_i^a}= \frac{\delta H_{\rm eff}}{\delta P_i^a} = P_i^a + D_i \lambda^a,
\label{h11}
\end{equation}
\begin{equation}
\dot{P_i^a}= -\frac{\delta H_{\rm eff}}{\delta Q_i^a} = (\nabla\times B)_i^a,
\label{h22}
\end{equation}
together with the constraint
\begin{equation}
\frac{\delta H_{\rm eff}}{\delta \lambda}=0 \Rightarrow D_i^2 \lambda^a + D_i P_i^a = -\frac{\partial U}{\partial \lambda^a}
\label{h33}
\end{equation}
are easily seen to be equivalent to (\ref{m11}, \ref{m22}).
Substituting (\ref{pq}) in (\ref{heff}), in order to express the Hamiltonian and the energy in terms of the physical fields,
one gets
\begin{equation}
H_{\rm eff} =\int_3 \frac{1}{2} E_i^a E_i^a + \frac{1}{2} B_i^a B_i^a - U.
\label{energy}
\end{equation}
As far as the bubble solution is concerned, its energy is
\begin{equation}
\epsilon_{\rm b}= \int_3 \frac{1}{2} (\nabla \lambda)^2 -U,
\label{eb}
\end{equation}
and is positive (much like the action of an instanton that mediates vacuum decay for a 
three-dimensional theory with an inverted potential). 
Here, dimensional arguments from (\ref{m0}, \ref{m4}) show that
$\epsilon_{\rm b} \approx \frac{\mu}{C_2 \alpha_s}$.

It is well-known that instanton solutions are unstable
against expansion, so it is important to check the problem of stability.
It is easy to see, however, that the bubble solutions, as well as all solutions to the equations of motion
stemming from (\ref{meff2}) or (\ref{heff}), are classically stable, since the second variation 
of the Hamiltonian (\ref{heff}) around a solution  is
\begin{equation}
\delta^2 H_{\rm eff} = \int_3 (\delta P_i^a)^2 + \delta\lambda^a (-D_i^2 - U'') \delta\lambda^a
                                  + \delta P_i^a D_i \delta \lambda^a
\end{equation}
(without the variation from the magnetic term, which is obviously positive).
The operator $-\nabla^2 - U''$ has a negative eigenvalue, corresponding to the 
similar tunneling problem mentioned before. Here, however, all variations have to satisfy the constraint
(\ref{h33}), hence
$(- D_i^2 - U'') \delta \lambda^a = D_i \delta P_i^a$,
and the solutions are stable, since then $\delta^2 H_{\rm eff} = \int_3 (\delta P_i^a)^2 \geq 0$ (plus terms from the variation of the 
canonical coordinates from the magnetic field, which are also positive).
In particular, the bubble solutions derived here are stable (although not topologically) soliton solutions, ``glueballs'' of the (chromo)-electric field.

\section{The Euclidean action}

The Euclidean action obtained from
(\ref{meff1}) is
\begin{equation}
S_{\rm E, eff}= \int \frac{1}{2} E_i^a E_i^a + \frac{1}{2} B_i^a B_i^a + 
                        \lambda^a D_i  E_i^a - U(\lambda).
\label{ee}
\end{equation}
The rotations involved are
\begin{eqnarray}
\nonumber
S_{\rm M} &\rightarrow& i S_{\rm E} \\ \nonumber
t &\rightarrow& - i \tau \\ \nonumber
A_0, \lambda, \partial_0 &\rightarrow& i A_0, i \lambda, i \partial_\tau,
\end{eqnarray}
integrations are over $d\tau d^3x$,
and the Euclidean equations are
\begin{equation}
\frac{\delta}{\delta\lambda^a}=0 \,\Rightarrow
D_i E_i^a = \frac{\partial U}{\partial \lambda^a},
\label{e1}
\end{equation}
\begin{equation}
\frac{\delta}{\delta A_0^a} =0 \,\Rightarrow
D_i^2 \lambda^a = D_i E_i^a,
\label{e2}
\end{equation}
\begin{equation}
\frac{\delta}{\delta A_i^a} =0 \Rightarrow
D_0 \, E_i^a =-  (D\times B)_i^a + D_0 D_i \lambda^a,
\label{e3}
\end{equation}
from which some preliminary observations can be made:

a) The configuration with $\lambda =0$ and the gauge fields also equal to zero, is a vacuum solution, $\Omega_0$, of both the
Euclidean and Minkowski equations, with zero energy.

b) The bubble solutions are not stationary points of the Euclidean equations of motion, so apart from their 
classical stability,  their quantum  stability against tunneling is also possible.

c) The Euclidean action has an imaginary part from the continuation of the effective potential, which grows for
large values of $\lambda$ and hints to instabilities. 

d) All points that satisfy $\lambda^2 =\mu^2$, with the gauge fields equal to zero, are gauge equivalent copies of the same vacuum, $\Omega_\mu$, 
a solution of both the Minkowski and the Euclidean equations, with positive Euclidean action and energy density.
Although it is a maximum of $-U$, it is classically stable by the previous discussion (as is $\Omega_0$).
Specifically, in the $A_0=0$ gauge, the vacuum $\Omega_\mu$ consists of time-independent, covariantly constant
configurations, $\lambda(\vec{x})= \omega(\vec{x}) \, \bar{\lambda}\, \omega(\vec{x})^{-1}$,
with $\bar{\lambda}$ a fixed adjoint vector with $\bar{\lambda}^2=\mu^2$ (and $F_{\mu\nu}=0$, with $A_i=\frac{i}{g}\omega(\vec{x}) \, \partial_i\, \omega(\vec{x})^{-1}$ as usual).

e) The Euclidean solutions generally satisfy $D_i^2 \lambda^a = \partial U / \partial \lambda^a$.
For  $E_i=\partial_i \lambda$ and $A_i, B_i=0$ this becomes
\begin{equation}
\nabla^2 \lambda^a = \frac{d^2 \lambda^a}{d r^2} + \frac{2}{r}\frac{d \lambda^a}{d r} =
\frac{\partial U}{\partial \lambda^a},
\label{et}
\end{equation}
which, if it were not for the ``friction'' term, would describe ``rolling'' of $\lambda$, from $\mu$, down through $0$, to the opposite but equivalent  point of $\Omega_\mu$.
It is similar to the equation for a three-dimensional soliton, which does not exist, because of Derrick's theorem \cite{cw}.
The possibility of solutions of the full Euclidean equations will be discussed later.

f) There is no solution of (\ref{et}) starting from $\lambda=0$ that would describe decay of the $\Omega_0$ vacuum
to larger values of $\lambda$ with negative energy density; this does not rule out, of course, other Euclidean solutions with more general field configurations.

g) At both $\Omega_0$ and $\Omega_\mu$, the remaining equations in Minkowski and Euclidean spacetime (\ref{m1}-\ref{m3})
and (\ref{e1}-\ref{e3}) are equivalent to the usual Yang-Mills equations, $D_\mu F^{a \mu\nu}=0$,
so one expects the well-known, Lorentz-invariant physics and results.

\section{Confinement}

Non-zero values of $\lambda$ signal confinement.
In fact, $\Omega_\mu$, with a constant $\lambda^2=\mu^2$ and the 
gauge fields equal to zero, is the confining vacuum. The diagram of Fig.~5, with external insertions of constant $\lambda$ (two smaller blobs),
gives, in the static limit, a factor of 
$C_2 \frac{g^2 \lambda^2}{\vec{k}^4}$ that corresponds to a confining potential between two colored sources (two larger blobs).
Because of the combinatorics and the Feynman rules described before, there are no other diagrams that contribute to this order in the static limit; there
are higher loop diagrams, but these do not spoil the result.

Thus, the confining interaction arises from the $\lambda$-condensate, with the same mechanism that generated
the effective potential term for $\lambda$.

Two charges that are a small distance apart in the perturbative vacuum have a Coulomb interaction (\ref{coulomb}).
As they move further apart, stable bubble solutions, with adjacent electric field dipoles, are formed between them.
Bubble condensation eventually drives the system to the confining vacuum.

As was explained before, the confining vacuum, $\Omega_\mu$, has higher energy density than the perturbative vacuum, $\Omega_0$, and
is classically stable (essentially because of kinetic and gradient terms). Its quantum mechanical stability depends on the Euclidean action and will be discussed in the following Section.

Obviously, the confining vacuum consists of the entire set of points $\lambda^2=\mu^2$, as described in comment d) of Section 4, and  there is no spontaneous symmetry breaking, since $\lambda$ is not a dynamical field with additional degrees of freedom or a kinetic term. 

\section{More on Euclidean solutions}

First, I will show that there are no finite action solutions of the Euclidean equations that mediate the decay of the confining vacuum, so, besides its classical stability, it is also quantum mechanically stable.
In fact, one is interested in the difference
\begin{equation}
{\cal B}=S_{\rm E, solution} -S_{\rm E, background},
\end{equation}
where, in our case, the background configuration is the confining vacuum with $\lambda^2 = \mu^2$, which decays via a solution to the Euclidean equations (with the boundary condition that it tends to $\mu^2$ at Euclidean infinity). From (\ref{ee}) we have
(integrations are over four-dimensional Euclidean spacetime)
\begin{eqnarray}
\nonumber
	{\cal B} &=& \int \, \frac{1}{2} E^2 + \frac{1}{2} B^2 + \lambda^a \,D_i E_i^a -(U-U(\mu^2))=  \\ 
 & =& I_E +I_B+I_C +I_U,
\end{eqnarray}
with obvious notation, and $I_E, I_B > 0$, $I_U <0$.

After a rescaling of the fields, with $A_M (x_M) \rightarrow \alpha A_M (\alpha \, x_M),\,\, \lambda(x_M) \rightarrow \lambda(\alpha \, x_M)$, we have
\begin{equation}
{\cal B}(\alpha)=I_E + I_B + \frac{1}{\alpha} I_C + \frac{1}{\alpha^4} I_U,
\end{equation}
and, if the field configuration is a solution of the Euclidean equations at $\alpha =1$, we get 
$I_C + 4 I_U =0$.  Here $x_M=x_0, x_i$ are the Euclidean coordinates ($x_0$ is the Euclidean time $\tau$), and after a different rescaling with $A_0 \rightarrow A_0 (x_0, \alpha x_i), \,\, A_i \rightarrow \alpha A_i (x_0, \alpha x_i),\,\, \lambda \rightarrow \lambda (x_0, \alpha x_i)$,
we have
\begin{equation}
{\cal B}(\alpha)=\frac{1}{\alpha}I_E + {\alpha}I_B + \frac{1}{\alpha} I_C + \frac{1}{\alpha^3} I_U,
\end{equation}
hence $I_E + I_C + 3 I_U = I_B$. Combining the last two results we get
$I_E = I_B + I_U$, and, in particular, $I_E < I_B$.

However, the two Euclidean equations (\ref{e2}) and (\ref{e3}), can be also written as
\begin{equation}
D_i (E_i^a - D_i \lambda^a )=0,
\label{e22}
\end{equation}
\begin{equation}
D_0 \,( E_i^a - D_i \lambda^a)=-  (D\times B)_i^a,
\label{e33}
\end{equation}
and can be compared to the usual Euclidean equations for Yang-Mills, which, assuming $O(3)$-symmetry \cite{taubes}, are solved by
\begin{equation}
E_i^a - D_i \lambda^a = \pm B_i^a.
\label{taubes1}
\end{equation}

(\ref{taubes1}) leads to
\begin{equation}
I_E =I_B + \int \frac{1}{2} (D_i \lambda)^2  \pm \partial_i (\lambda^a \, B_i^a),
\label{taubes2}
\end{equation}
hence, at least for field configurations that fall sufficiently fast and are topologically trivial at infinity, $I_E > I_B$, in contrast with the previous result.

 Thus, there are no Euclidean solutions with finite ${\cal B}$ that mediate the decay of the confining vacuum.

In the previous derivation, the first Bianchi identity,
\begin{equation}
D_i B_i^a =0,
\label{bianchi1}
\end{equation}
was used,
and the second Bianchi identity,
\begin{equation}
D_0 B_i^a = - (D \times E)_i^a,
\label{bianchi2}
\end{equation}
is seen to hold also for the reshuffling of $E_i - D_i \lambda$ in (\ref{e2}, \ref{e3}).
It is in (\ref{e1}) or (\ref{et}) that the new non-linearities are expressed;
(\ref{e2}, \ref{e3}) can be treated as the usual Yang-Mills and generally lead to (\ref{taubes1}).

Next, I will show that there are no solutions of the Euclidean equations that mediate the decay of the perturbative vacuum. Since $U(\lambda^2=0)=0$, I will show that there are no finite action solutions of the Euclidean equations whatsoever.

The derivation proceeds as before, except that now, the related $I_U = \int -U $ is not positive or negative definite. Still, the relations $I_E = I_B +I_U$, $I_E +I_C +3 I_U=I_B$,
and $I_C+4 I_U=0$
are derived,
and (\ref{taubes1}) holds for every Euclidean solution with at least $O(3)$-symmetry (which will be assumed). Then (\ref{taubes2}), for sufficiently smooth and topologically trivial solutions, leads to
\begin{equation}
I_E =I_B + \int -\frac{1}{2} \lambda\, D^2_i \lambda =
                I_B -\frac{1}{2} I_C 
\label{taubes3},
\end{equation}
after using (\ref{e2}).

The previous relations can only be consistent if $I_U = I_C =0$.
Now $I_C = \int \lambda^a \frac{\partial U}{\partial \lambda^a}$ also holds on the Euclidean solutions by (\ref{e1}, \ref{e2}),
and, for a Coleman-Weinberg potential of the form appearing here,
$U= c \, \lambda^4 ( \ln \frac{\lambda^2}{\mu^2} -\frac{1}{2})$, with $c$ a constant, and
$\lambda^2 = \lambda^a \lambda^a$,
we have
\begin{equation}
 \lambda^a \frac{\partial U}{\partial \lambda^a}  = 4 U + 2 c \, \lambda^4 ,
\end{equation}
hence $I_C = 4 I_U + 2c\int \lambda^4$, and
the previous relations cannot be satisfied for a non-trivial (non-zero $\lambda$) Euclidean solution with finite action.

The only solutions to the Euclidean equations have constant $\lambda=0$ (or $\lambda^2=\mu^2$, in the previous case) and are the usual Yang-Mills instantons, with $\vec{E}^a=\pm \vec{B}^a$ and $I_E=I_B$.

For the field configurations at infinity, it should be noted that,
generally, there is no spontaneous symmetry breaking,
and there is no $U(1)$-electromagnetic charge.
This, of course, does not forbid solutions that may have an overall $U(1)$ symmetry.
As far as vacuum transitions and decays are concerned, however, it is difficult to see how they could involve a net ``magnetic'' charge.

Topologically non-trivial solutions to generalizations of the Euclidean equations (\ref{e1}, \ref{e2}, \ref{e3})
or (\ref{e1}, \ref{taubes1}) in other backgrounds may exist. Then, their 
asymptotic values, with $\lambda^2$ going to $\mu^2$ at spatial infinity, may correspond to a non-trivial second homotopy group,
but they are not necessarily  topologically stable. Their ``core'' is in the perturbative vacuum, $\lambda=0$, which has lower 
energy and lower Euclidean action (at least in the $-U$ term). Their instabilities reflect the fact that they are possible Euclidean solutions, describing quantum tunnelling phenomena.

In fact, when one considers thermal fluctuations and matter fields, it is expected that
such solutions to the Euclidean field equations, with finite ${\cal B}$, and trivial or non-trivial topology, exist
(a generalization of the ansatz of \cite{witten} may be relevant).
At finite temperature, there are modifications to the generated effective potential; the fact that the thermal phase transition in Coleman-Weinberg models
is of the first order, already gives a prediction for the order of the deconfining phase transition. At finite temperature, however, the situation
is intrinsically non-covariant because of the plasma background, and there are more terms generated in the effective action that need to be considered.

Colored, fermionic, matter fields also couple to $\lambda$ via Gauss's law, 
as in (\ref{l1}) and its generalization to the non-Abelian case,
and can ``tilt'' the effective potential, $U(\lambda)$, thereby enabling
finite action solutions of (\ref{et}) and (\ref{taubes1}) that connect the two vacua.

\section{Energy-momentum tensor, global current, Lorentz invariance, the bag model, chiral symmetry breaking, etc.}

Because of the effective potential term, generated by quantum effects,
the effective Lagrangian 
\begin{equation}
{\cal L_{\rm eff}} = \frac{1}{2} E_i^a\,E_i^a - \frac{1}{2} B_i^a \, B_i^a +\lambda^a \, D_i E_i^a +U(\lambda)
\end{equation}
singles out Gauss's law and the time component that it involves. Eventually, however,
two vacua emerged. The perturbative, Coulomb vacuum, with $\lambda^2 =0$ and $U=0$, where
one has the usual perturbative theory, and the confining vacuum, with $\lambda^2 = \mu^2$ and positive $-U$. They are both classically and quantum mechanically stable, and one may further examine the symmetry properties of the theory, first by considering the general expression,
\begin{equation}
T^{\mu \nu}= \sum_n \frac{\partial {\cal L_{\rm eff}}}{\partial(\partial_\mu \phi_n)}
                                   \,\partial^\nu \phi_n
                      - \eta^{\mu\nu} {\cal L_{\rm eff}},
\end{equation}
where $\phi_n$ denotes all the fields and their components in ${\cal L}$.
This
can be improved with the addition of 
\begin{equation}
\Delta T^{\mu\nu} = -\partial_\rho (F^{a \mu\rho}A^{a \nu}),
\end{equation}
a total derivative with $\partial_{\mu} \Delta T^{\mu\nu} =0$,
and the further addition of
$\Delta {\tilde T^{\mu\nu}}$ with
\begin{equation}
\Delta {\tilde T_{00}}=\partial_i (D_i \lambda^a \,A_0^a),\,\,
\Delta {\tilde T_{i0}}=\partial_0 (D_i \lambda^a \,A_0^a),
\end{equation}
\begin{equation}
\Delta {\tilde T_{ij}}=\partial_0 (D_i \lambda^a \,A_j^a),\,\,
\Delta {\tilde T_{0j}}=\partial_i (D_i \lambda^a \,A_j^a),
\end{equation}
also a total derivative with $\partial_{\mu} \Delta {\tilde T^{\mu\nu}} =0$.
The final expression
\begin{equation}
\Theta^{\mu\nu}=T^{\mu\nu}+\Delta T^{\mu\nu} +\Delta {\tilde T^{\mu\nu}}
\end{equation}
also satisfies $\partial_\mu \Theta^{\mu\nu} =0$, and
\begin{equation}
\Theta^{00}=\frac{1}{2}E_i^a E_i^a +\frac{1}{2}B_i^a B_i^a -U ={\cal H}_{\rm eff},
\end{equation}
coincides with the expression in (\ref{energy})
(modulo surface terms of the form $\lambda^a D_i E^a_i + E^a_i D_i\lambda^a $ that will be ignored for fields that fall sufficiently fast and are topological trivial at infinity).

However, the remaining components of $\Theta^{\mu\nu}$, for a general
$D_i \lambda^a \neq 0$, are neither symmetric nor do they form a Lorentz-invariant tensor (for example $\Theta^{0i}=(\vec{E}-\vec{D\lambda})\times \vec{B}$,
$\Theta^{i0}=\vec{E}\times \vec{B}$).

At the vacua,  where $D_i\lambda^a =0$,
\begin{equation}
\Theta^{\mu\nu}=\Theta^{\mu\nu}_{Y-M} - U \, \eta^{\mu\nu}
\end{equation}
is the symmetric, conserved, Lorentz invariant, energy-momentum tensor of the theory.

$\Theta^{\mu\nu}_{Y-M}$ is the usual, traceless, energy-momentum tensor
for perturbative Yang-Mills ($\Theta^{\mu\nu}_{Y-M}=F^{a \mu \rho}{F^{a \nu}}_\rho -\frac{1}{4} \eta^{\mu\nu}F^{a\lambda\rho}F^a_{\lambda\rho})$
 and coincides with $\Theta^{\mu\nu}$ at the perturbative vacuum, where $U=0$. At the confining vacuum, with $\lambda^2 = \mu^2$ one 
also gets a Lorentz invariant theory, with the additional energy-momentum tensor of
a bag model, $-U\, \eta^{\mu\nu}$, with a positive constant, $-U(\mu^2)$,
and a positive, non-zero trace $\Theta^\mu_\mu = - 4 U(\mu^2)$.

Physics is Lorentz invariant at both vacua. However, the transitions between the two vacua involve non-trivial (chromo)-electric and -magnetic field configurations (for example, the solitons, ``glueballs'' of finite radius of the chromoelectric field, described earlier) and cannot be decribed in a Lorentz invariant manner.
Obviously, a physical quantity, like the rate of a transition from one vacuum to the other, should be Lorentz invariant.

 However,  for pure Yang-Mills theory, the two vacua do not decay.
In fact, the lack of a Lorentz invariant energy-momentum tensor that connects them is related to that.
The vacua may ``jump'' to configurations where $-U$ is negative, with fluctuations of $\lambda$ of order $\mu$, from where, depending on the background involved, they may settle to the absolutely stable confining vacuum via processes like the condensation of the previously derived solitons (``glueballs''). But there is no solution of the Euclidean equations, for pure Yang-Mills, that describes such a process.
Such fluctuations and transitions between the two vacua obviously happen in the presence
of background configurations with finite temperature and/or fermion density
(or other fluctuations, like during the purely academic example where one considers two initially static, colored sources and pulls them apart)
but there, the issue of Lorentz invariance, and especially its description via an effective action, is more involved, or even moot.

Continuing the investigation of the symmetry properties of the theory,
one may consider the expression
\begin{equation}
J_\mu^a = \sum_n \frac{\partial {\cal L_{\rm eff}}}{\partial(\partial_\mu \phi_n)}
\frac{\delta \phi_n}{\delta \alpha^a},
\end{equation}
where the $\alpha^a$'s are constant, global parameters of the gauge group.
Again, for general configurations with $D_i \lambda^a \neq 0$, this expression contains factors of $D_i \lambda^a$ and is not Lorentz invariant.

At both vacua, where $D_i \lambda^a =0$, this expression is Lorentz invariant, it is the same at both vacua, and coincides with the usual expression for the Yang-Mills theory
($J_\mu^a= f^{abc} A^{b \rho} F^c_{\rho\mu}$).
It is a conserved and Lorentz invariant current, but it is not gauge invariant. The reason is the same at both vacua. Namely, even in the confining vacuum, the theory still contains the massless gluons and the Coulomb interaction, as in Fig.~1, in addition to the confining interaction of Fig.~5. The limitations of \cite{ww}, therefore, continue to hold
(it is noted again that, for both $T_{\mu\nu}$ and $J_\mu$, the calculations and improvements were made for fields that fall sufficiently fast and are topologically trivial at infinity).

Finally, although this work investigates the vacuum structure of pure Yang-Mills theory, it is also of interest to make another connection with older phenomenological models and notice that, when fermionic, matter fields, are included, they also couple to $\lambda$ via Gauss's law, as in (\ref{l1}) and its generalization to the non-Abelian theory.
In the confining vacuum, therefore, the $\lambda^2 = \mu^2$ condensate gives an effective interaction of the Nambu--Jona-Lasinio type,
\begin{equation}
G_{NJL}  (\bar{\psi} \gamma^0 \psi)^2.
\end{equation}
In terms of the infrared and ultraviolet cut-offs, $\Lambda_{IR}, \Lambda_{UV}$, needed to define the effective interaction \cite{chiral},
$G_{NJL}\approx \frac{g^2 \, \mu^2}{\Lambda_{IR}^4}$, and chiral symmetry breaking occurs when $G_{NJL} \stackrel{>}{\sim}  \frac{1}{\Lambda_{UV}^2}$,
and is also related to the confining mechanism, as expected.

\section{Comments on the strong-CP problem}

It is well-known that, in the $A_0 =0$ gauge, space-dependent gauge transformations
in the perturbative vacuum
 fall into topologically distinct
sectors, $|n>$, that correspond to the winding-number, $n$, of maps of the compactified three-dimensional space to 
the group, $G$. These then combine into the so-called $\theta$-vacuum,
$|\theta> = \sum_n e^{-i n \theta}|n>$, and $\theta$ is equivalent to an additional periodic parameter multiplying
a CP-violating term, proportional to $\vec{E}\cdot\vec{B}$, in the action \cite{cw}.

The same situation occurs in the confining vacuum.
It also has a similar, non-trivial structure, described in comment d) of Section 4, and
the usual Yang-Mills instantons exist there too, as can be easily seen from the Euclidean equations
(\ref{e1}-\ref{e3}) or (\ref{taubes1}) and the discussion of Section~6.
The two vacua for pure Yang-Mills are completely stable, but the presence of two $\theta$-vacua, that can be connected via a multitude of physical processes,
makes it very difficult, if not impossible, for $\theta$ to have any value other than zero.

For example, the stable soliton ``glueball'' solutions derived in Section 3
are topologically trivial configurations in the perturbative vacuum, with fixed directions in color space;
one may imagine, however, their condensation in three-dimensional space with
varying such directions, 
so that they cover it, for example, with winding number unity, thereby
corresponding to a topology-changing transition, 
$|n>\rightarrow |n+1>$, from $\Omega_0$ to $\Omega_\mu$.

Then, since
$|\theta>\rightarrow\sum_n e^{-i n \theta}|n+1> = e^{i \theta} |\theta>$,
consistency implies that $\theta$ is limited to values essentially zero,
and the $\theta$-vacuum to an unglorious demise.
The fact that the usual Yang-Mills instantons (solutions of $\vec{E}^a=\pm \vec{B}^a $), as well as  all the
Euclidean and Minkowski solutions of the usual Yang-Mills equations, still exist in 
both the perturbative and the confining vacuum, as was repeatedly mentioned before, ensures that most of the traditional folklore
regarding, for example, the solution of the $U(1)$-problem, remains unharmed.

Both vacua, for pure Yang-Mills at zero temperature, 
are completely stable.
Physically, however, the two vacua are obviously  connected, as was discussed before,
and transitions between them, for example, in backgrounds of finite temperature or density, are expected to exist.
Unless every transition between the two vacua follows the $|n>\rightarrow |n>$ pattern, $\theta$ is effectively zero.
The solution to the strong-CP problem, therefore, is also related to the confining vacuum and mechanism, as 
has been argued before \cite{cp}.

\section{Discussion}

In the usual process of quantization of a non-Abelian gauge theory, it is often stated that $A_0$ acts as a Lagrange 
multiplier enforcing Gauss's law (a constraint equation that includes $A_0$, which also poses as the associated Lagrange multiplier). Sometimes a shift $A_0 + \lambda \rightarrow A_0$  is performed, one generally moves between the Hamiltonian and Lagrangian formalisms, and finally one arrives at a set of Feynman rules
that claim to express the constraint, but could have been derived anyway, by simply inserting unity
in the path integral and splitting it via the Fadeev-Popov trick, without mentioning constraint quantization whatsoever.

It is the claim of this work that the afore-mentioned shift and general procedure of treating the constraint, are not exact beyond tree-level,
and that there are some ``left-over'', $\lambda^2$-dependent, terms described here. Eventually, by treating these terms, I demonstrated the mass-gap of the Yang-Mills theory,
clarified the structure of the confining vacuum and confining mechanism, their relation to the perturbative vacuum, and gave some quite suggestive arguments for the subsequent resolution of the strong-CP problem.

The Feynman rules used here came from a procedure that is not initially explicitly Lorentz invariant, but they combine to reproduce all known perturbative processes. 
All classical solutions of the Lorentzian equations of motion, were  shown to be classically stable, and both vacua were also shown to be quantum mechanically stable.
At both vacua, the remaining equations of motion are the usual ones, $D_\mu F^{a\mu\nu}=0$.
The energy-momentum tensor is the usual, Lorentz invariant, expression at both vacua,  with the addition of a Lorentz invariant, bag model, contribution.

As was expected, the confining vacuum provides the resolution and explanation of the strong-CP problem, as well as chiral symmetry breaking, and enjoys a Lorentz invariant energy-momentum tensor, with a non-zero trace, that contributes to the baryon mass, and may have other interesting experimental and cosmological consequences.
The usual, perturbative results, like, for example, the trace anomaly, still exist, but there is no need to invoke dubious gauge, or fermion, field ``condensates'', although, of course, additional, ``non-perturbative'' effects may also exist.

Euclidean or Lorentzian solutions that connect the two vacua, like
the stable (Lorentzian) soliton solutions derived here (spherically symmetric ``glueballs'' of the chromo-electric field) obviously cannot be manifestly  Lorentz-frame independent.
The two vacua, for pure Yang-Mills at zero temperature, are absolutely stable and transitions between them can only occur via fluctuations (of order $\mu$) such as, for example, in a background of finite temperature or density. Then, the problem of Lorentz invariance of the relevant quantities (if it can be posed) becomes more involved and beyond the scope of the effective action derived and used here. As far as pure Yang-Mills at zero temperature is concerned, the two vacua are not connected, and physics is Lorentz invariant at each one. 

Although many expected properties were derived with the present approach (mass gap, confinement, chiral symmetry breaking, resolution of the strong-CP  problem, bag constant) the possibility of a quantum field theory (especially the ubiquitous Yang-Mills theory) with two separate, stable, Lorentz invariant vacua is quite surprising and far reaching in terms of the search of a unified theory, the applicability and generality of the Lagrangian formalism, the effective field theory approach, and other problems beyond the strong interactions.
The scales involved in quantum chromodynamics, as well as properties such as factorization \cite{al}, make it difficult to investigate the characteristics of the two vacua. For other non-Abelian gauge theories that may exist at higher energy scales, this may not be the case. The experimental and cosmological consequences of processes at these scales then will be quite distinct.

Also, the relation between different phases like the confining, Coulomb, and the spontaneously broken, Higgs phase of a non-Abelian gauge theory can hopefully be investigated in future works, using extensions of the present formalism. In the weak interactions, the scales involved ensure that they are typically observed in the broken phase. Generally, for a non-Abelian gauge theory, there is an interplay between its characteristic generated scale (the scale $\mu$, derived in this work), the scales of the masses of the fermionic matter fields that are subject to this interaction, any ``Higgs''-type scalar fields that are involved, as well as the ``environment'' with the temperature and other experimental scales, leading to a rich structure of different phases (confining, Coulomb, and Higgs, among them) and phase transitions between them, that can hopefully be more thoroughly investigated using some of the tools described in this work.

\newpage

\begin{figure}
\centering
\includegraphics[width=90mm]{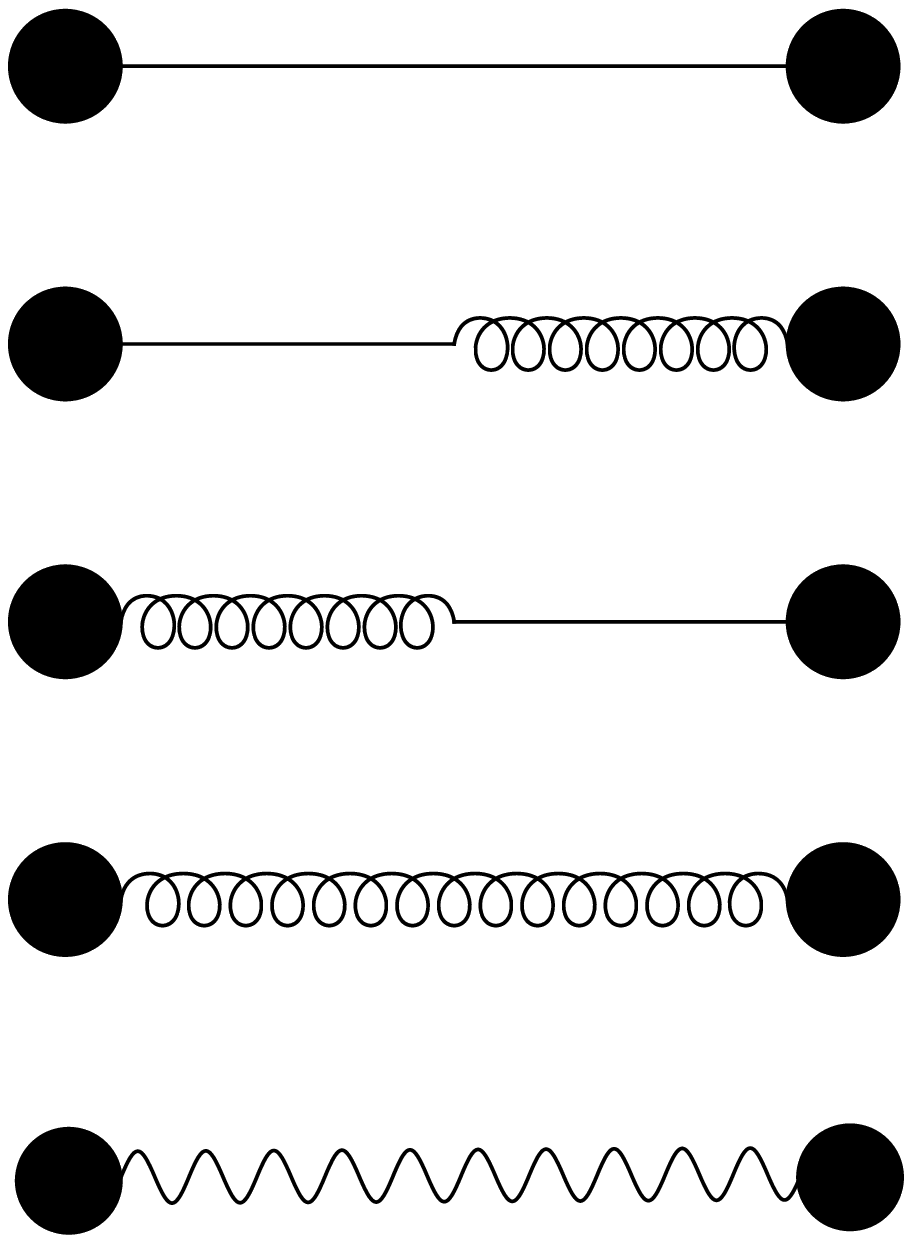}
\caption{The propagators  combine to reproduce the Coulomb interaction between two
static sources (large blobs).
Solid, wavy and curly
lines denote the $A_0$, $A_i$ and $\lambda$ fields respectively.}
\end{figure}

\begin{figure}
\centering
\includegraphics[width=90mm]{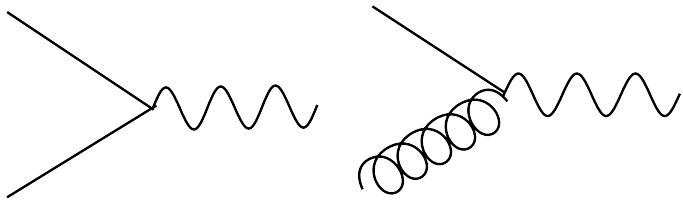}
\caption{Two vertices for the non-Abelian theory.}
\end{figure}

\begin{figure}
\centering
\includegraphics[width=90mm]{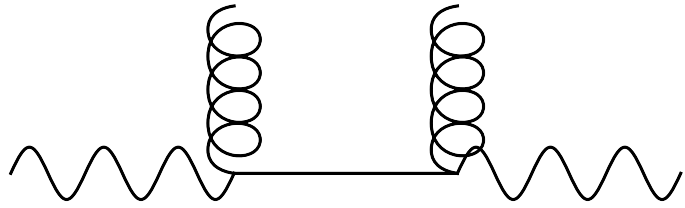}
\caption{The modifications to the i-j propagator.}
\end{figure}

\begin{figure}
\centering
\includegraphics[width=90mm]{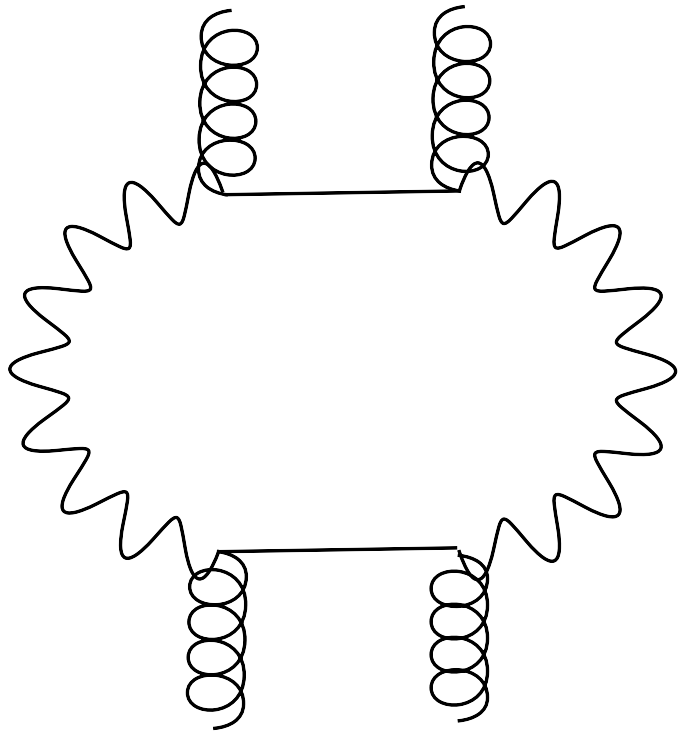}
\caption{A diagram for the generated effective potential.}
\end{figure}

\begin{figure}
\centering
\includegraphics[width=90mm]{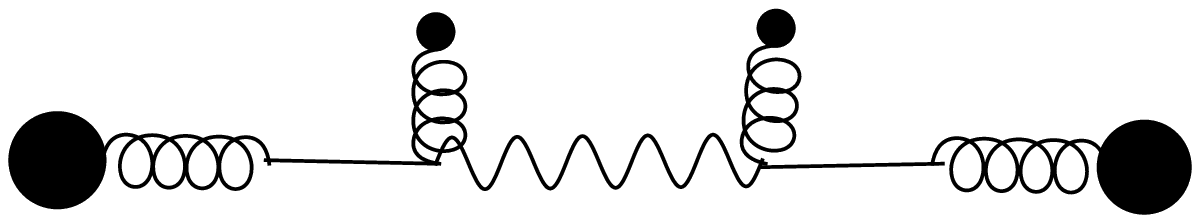}
\caption{The diagram for the confining interaction.}
\end{figure}


\begin{thebibliography}{99}

\bibitem{dkm} D.~Metaxas, {\it Phys. Rev.} {\bf D75}, 065023 (2007).

\bibitem{CW}S.~Coleman and E.~J.~Weinberg, {\it Phys. Rev.} {\bf
D7}, 1888 (1973).

\bibitem{kogut} J.~Kogut and L.~Susskind, {\it Phys. Rev.} {\bf D9}, 3501 (1974).

                        M.~Bander and P.~Thomas, {\it Phys. rev.} {\bf D12}, 1798 (1975).

\bibitem{conf}   G.~'t~Hooft, {\it Nucl. Phys. Proc. Suppl.} {\bf 121}, 333 (2003).

                        G.~B.~West, {\it Phys. Lett.} {\bf B115}, 468 (1982).

                         D.~Dudal, {\it Phys. Lett.} {\bf B677}, 203 (2009).

                         A.~V.~Zayakin and J.~Rafelski {\it Phys. Rev.} {\bf D80}, 034024 (2009).

\bibitem{bag}  P.~Hasenfratz, {\it Phys. Rept.} {\bf 40}, 75 (1978).

                      T.~D.~Lee, {\it Phys. Rev.} {\bf D19}, 1802 (1979).

                      H.~Reinhardt, {\it Phys. Rev. Lett.} {\bf 101} 061602 (2008).

                      M.~J.~Neubelt, A.~Sampino, J.~Hudson, K.~Tezgin and P.~Schweitzer, {\it Phys. Rev.} 
                                                                {\bf D101}, 034013 (2020).
              
\bibitem{chiral} U.~Vogl and W.Weise, {\it Prog. Part. Nucl. Phys.} {\bf 27}, 195 (1991).

                       S.~P.~Klevansky, {\it Rev. Mod. Phys.} {\bf 64}, 649 (1992).

                       D.~Nickel, {\it Phys. Rev.} {\bf D80}, 074025 (2009).

                       R.~S.~Hayano and T.~Hatsuda, {\it Rev. Mod. Phys.} {\bf 82}, 2949 (2010).

\bibitem{cp} M.~A.~Shifman, A.~I.~Vainshtein and V.~I.~Sakharov, {\it Nucl. Phys.} {\bf B166}, 493 (1980).

                   H.-Y.~Cheng, {\it Phys Rept} {\bf 158}, 1 (1988).

                     T.~Ibrahim and P.~Nath, {\it Rev. Mod. Phys.} {\bf 80}, 577 (2008).

\bibitem{cw} S.~Coleman, {\it Aspects of Symmetry}, Cambridge Univ. Press (1985).

E.~J.~Weinberg, {\it Classical solutions in quantum field theory}, Cambridge Univ. Press (2012).

\bibitem{taubes} C.~H.~Taubes, {\it Commun. Math. Phys.} {\bf 75}, 207 (1980).

\bibitem{witten} E.~Witten, {\it Phys. Rev. Lett.} {\bf 38}, 121 (1977).

\bibitem{ww} S.~Weinberg and E.~Witten, {\it Phys. Lett.} {\bf B96}, 59 (1980).

\bibitem{al}  N.~H.~Christ, B.~Hasslacher, and A.~H.~Mueller, {\it Phys. Rev.} {\bf D6}, 3543 (1972).

                   W.~A.~Bardeen, A.~J.~Buras, D.~W.~Duke and T.~Muta, {\it Phys. Rev.} {\bf D18} 3998 (1978).

                   A.~H.~Mueller, {\it Phys. Rept.} {\bf 73}, 237 (1981).

                  J.~C.~Collins, D.~E.~Soper and G.~F.~Sterman, {\it Adv.Ser.Direct.High Energy Phys.} {\bf 5}, 1 (1989).

\end{thebibliography}
\end{document}